\documentstyle[aps,pre,preprint,epsf,epsfig]{revtex}
\begin{document}
\draft
%
%

\title{Numerical estimation of entropy loss on dimerization:
improved prediction of the quaternary structure of the GCN4 leucine
zipper}

\author{Jorge Vi\~nals$^{1}$, Andrzej Kolinski$^{1,2}$, and 
Jeffrey Skolnick$^{1}$}
\address{$^{1}$ Laboratory of Computational Genomics, Donald Danforth Plant
Science Center, 975 North Warson Road, St. Louis, MO 63132, $^{2}$
Department of Chemistry, University of Warsaw, 02-093 Warsaw, Poland}

\date{\today}
\maketitle

\begin{abstract}
A lattice based model of a protein is used to study the dimerization
equilibrium of the GCN4 leucine zipper. Replica exchange Monte Carlo
is used to determine the free energy of both the monomeric and dimeric
forms as a function of temperature. The method of coincidences is then
introduced to explicitly calculate the entropy loss associated with
dimerization, and from it the free energy difference between
monomer and dimer, as well as the corresponding equilibrium reaction 
constant. We find that the entropy loss of dimerization is a strong 
function of energy (or temperature), and that it is much
larger than previously estimated, especially for high energy states.
The results confirm that it is possible to study the dimerization
equilibrium of GCN4 at physiological concentrations within the
reduced representation of the protein employed.
\end{abstract}

\pacs{}
 

\section{Introduction}

Reduced models of a protein have been shown to provide a possible route
for the estimation of the free energy of dimerization of relatively
short coiled-coils \cite{re:vieth95,re:vieth96,re:mohanty99}.
A key step in the calculation of the free energy
of dimerization concerns the entropy loss upon bringing two monomer
chains together to form the dimer. A
practical method for the numerical estimation of
this entropy loss by Monte Carlo simulation is discussed in this
paper.

The focus of our work is on the calculation of free energies of
dimerization, and in particular a re-analysis (using a subsequently
improved model) of prior research about the
folding thermodynamics of the GCN4 leucine zipper 
\cite{re:vieth96,re:mohanty99}. Leucine zippers belong to the class
of structural motifs that are known as coiled coils. Generically, they
comprise right handed $\alpha$ helices wrapped around each other with a
small left-handed super-helical twist \cite{re:crick53}. 
While leucine zippers can exist in both monomeric or dimeric form
\cite{re:oshea91}, GCN4 forms a dimer in the crystalline phase
\cite{re:oshea91}.

Numerical calculations of the 
free energy difference between the monomeric and dimeric forms of GCN4 have
already been given in \cite{re:vieth96,re:mohanty99}. In 
ref. \cite{re:vieth96}, 
the free energy of the monomer was computed by transfer matrix
methods, whereas the entropy change of dimerization was estimated by
Monte Carlo methods. Two monomers were placed in a parallel
configuration, and the configurational partition function was estimated
by placing the two monomers in registry with one another, but
considering different position of the relative starting point of each
segment. The Entropy Sampling Monte Carlo method of
Hao and Scheraga \cite{re:hao94} was used in ref. \cite{re:mohanty99}
to sample the configurational
space of both monomer and dimer. By using the values of the various
terms that contribute to the entropy loss of dimerization given in
\cite{re:vieth96}, the equilibrium dimer fraction was computed as a
function of temperature and monomer concentration. As a byproduct of the
calculation, other equilibrium quantities such as the helical content of
the monomer and dimer were obtained, as well as an analysis of the
existence of possible folding intermediaries.

In our present work, we use the Replica Exchange Monte Carlo method
\cite{re:swendsen86,re:geyer92,re:hukushima96} to obtain the canonical
distribution of both monomer and dimer forms separately. Re-weighting methods 
are then used to calculate the respective free energies as a function
of temperature
\cite{re:ferrenberg89a,re:ferrenberg89b}. In the final step, a method 
is described to place
both free energies on the same relative scale so that the free energy
difference between the monomer and dimer forms can be computed. We
believe that the method described in this paper allows a more accurate 
determination of the entropy loss of dimerization than previously
reported, and therefore it allows a
more accurate determination of the free energy difference between the
monomeric and dimeric forms as a function of temperature and
concentration.

\section{The protein model}

Our analysis is based on a a reduced model in which a protein 
is represented by a sequence of virtual bonds connecting effective particles
\cite{re:kolinski94,re:kolinski96}.
Each of these particles is assumed to be located at the center of mass of
the side chain and backbone $\alpha$ carbon. The
effective particles are embedded in a regular cubic lattice of fixed
spacing that allows for a fairly accurate representation of the backbone
of known protein structures.
This geometric part of the model has been checked against all
structures contained in the protein data bank
\cite{re:bernstein77}. The observed root mean squared deviation
between the lattice representation of any protein and its resolved
structure is typically below 0.8~\AA \cite{re:kolinski99}.
The actual resolution of the model is of course lower, typically of the
order of 2~\AA for small proteins.

Effective interactions (force fields) are introduced among the particles
that include generic (sequence independent), and sequence specific
contributions. The potentials associated with the generic type of interactions
are defined so as to introduce a bias toward
reasonable secondary structures. One such potential is introduced to
account for the fact that proteins exhibit a characteristic bimodal
distribution of neighbor residue distances, specially between the i-th
and (i+4)-th residues. Configurations corresponding to the larger
distance in the distribution are associated with proteins that exhibit
either $\beta$-type or expanded coils, whereas the shorter distance 
corresponds to helices and turns
\cite{re:kolinski99}. A second generic interaction further introduces
a bias toward certain packing structures such as helices and
$\beta$-type states. The first potential produces the required 
stiffness of the polypeptide chain, whereas the second provides for local 
cooperative motion during packing.

Sequence specific interactions are of three types, and include short
range interactions, long range, pairwise interactions, and many body
interactions. The short range pairwise potentials are of statistical
origin and are fitted to nonhomologous reference structures. This is 
accomplished by considering the known
distances between pairs of aminoacids that are separated by one through
four bonds along the chain. Chirality is also introduced by considering
an interaction between the i-th and (i+3)-th and (i+6)-th bonds to produce
the correct pitch. Long range interactions are also of statistical
origin, and are assumed to depend not only on relative distances between
the effective atoms, but also on the relative orientation between the
corresponding bonds (for example, residues of opposite charges are
attractive when the corresponding bonds are parallel to each other,
whereas the interaction is weak or repulsive when they are
anti-parallel).

Finally, although multi-body interactions are implicitly included (as an
unknown contribution) in the pairwise potentials determined from
inter-residue distances and bond angles, two additional terms are added
to model hydrophobic interactions and the known probability of a residue to
have a given number of parallel and anti-parallel contacts. The hydrophobic
potential is
estimated from the surface exposure of a given side chain, i.e., of all
possible contacts of a side chain, those that are not effectively
occupied by contacts with neighboring chains. The second
multi-body potential introduces a bias toward known propensities of
various aminoacids to pack their side chains in parallel or anti-parallel 
orientation.
Recent applications of the methodology include the improvement of 
threading based structure prediction \cite{re:kolinski99}, and direct
ab-initio folding studies \cite{re:kolinski00}.

\section{Monomer-dimer equilibrium and Monte Carlo method}
\label{sec:model}

At constant temperature and
in thermodynamic equilibrium, the concentration of freely 
associating monomers and dissociating dimers 
is governed by the reaction constant,
$K = [D]/[M]^{2}$, where $[M]$ and $[D]$ are the concentrations
of monomers and dimers respectively.
The mol fraction equilibrium constant can be expressed in terms of the 
canonical partition functions of both monomer and dimer as
\cite{re:mayer63},
\begin{equation}
\label{eq:kxdef}
K_{x} = c_{0} K = N \frac{Z_{D}}{Z_{M}^{2}} = N \frac{Q_{D}}{\sigma_{D}
Q_{M}^{2}},
\end{equation}
where $N$ is the total number of chains (i.e., two chains per dimer), 
$c_{0} = N/V$ the chain concentration in a system of volume $V$, and
$Z_{M}$ and $Z_{D}$ the canonical partition functions of the monomer
and dimer. The momenta 
degrees of freedom of both
monomer and dimer can be integrated out from their partition
functions, and the respective kinetic contributions cancel. We have therefore 
introduced the 
configurational partition functions $Q_{M}$ and $Q_{D}$, as well as
the symmetry factor $\sigma_{D}$ that takes into account the 
indistinguishability
of the two chains that form the dimer ($\sigma_{D} = 2!$ in the present
case) \cite{re:mayer63}. The purpose of the present paper is 
an estimation of $Q_{M}$ and $Q_{D}$ by the Monte Carlo method. 

Although it is in principle possible in to estimate the ratio of 
configurational partition functions in Eq. (\ref{eq:kxdef})
by direct simulation involving coexisting
monomers and dimers that transform into each other by association
or dissociation, we have found that this is not
practical. First, proteins are large molecules (even in the reduced
representation used in our work and described in Appendix \ref{ap:sicho}),
and only a small number of them can be placed in a computational
cell. Second, the protein models employed lack very long range interactions,
and hence there are large entropic contributions to the free energies which
are difficult to sample accurately over the large configurational space
of a protein.
We instead follow the approach of refs. \cite{re:vieth96,re:mohanty99} which
consists of two steps: an independent computation of the free energies of 
the monomer
and dimer forms, followed by a transformation to a common reference state so 
that the free energy difference between the two can be estimated.

The configurational partition functions $Q_{M}$ and 
$Q_{D}$ are estimated by the Replica
Exchange Monte Carlo method \cite{re:swendsen86}. Either a single
monomer or a single dimer are placed
in the computational cell, and a set of canonical Monte Carlo
simulations are performed at set of prescribed neighboring temperatures.
By conducting simulations involving only one monomer or one dimer
we are explicitly assuming that at physiological 
concentrations monomer-dimer or dimer-dimer interactions can be
neglected.
The simulation to obtain $Q_{D}$ involves two identical monomers constrained
during the course of the simulation to states in which there is at least
one inter chain contact.

We consider a set of $r$ independent canonical simulations conducted
in parallel at a set of neighboring inverse temperatures $\beta_{i}, 
(i=1, \ldots, r)$.
At fixed intervals during the course of the simulation, two configurations at 
different temperatures
(\lq\lq replicas'') are chosen at random, and their respective temperatures
exchanged with probability defined so as to preserve detailed balance as given
by the canonical probability distribution. Additional details about the
so-called Replica Exchange Monte Carlo method can be found in ref. 
\cite{re:swendsen86}. At each inverse temperature $\beta_{i}$, a sample of
$n_{i}$ statistically independent configurations is collected, and
the corresponding energy histograms $h_{i}(E)$ calculated with some
arbitrary energy binning $\Delta E$. Re-weighting
is then used to estimate the partition function by simultaneously solving
\cite{re:ferrenberg89a},
\begin{eqnarray}
\label{eq:probability_def}
p(E,\beta) & = & \frac{ \left( \sum_{i=1}^{r} h_{i}(E) \right) e^{- \beta
E}}{\sum_{i=1}^{r} n_{i} e^{-\beta_{i}E + f_{i}}}, \\
\label{eq:freeenergy_def}
e^{-f_{i}} & = & \Delta E \sum_{E} p(E,\beta_{i}),
\end{eqnarray}
where $f_{i} = \beta_{i} F(\beta_{i})$ is the dimensionless 
thermodynamic free energy at inverse temperature $\beta_{i}$. As is 
standard, free energy and canonical partition function are related through
$f_{i} = - \ln Q(\beta_{i})$. Also, for latter reference, we note that
the configurational density of states $W(E)$ is given by,
$$
W(E) = p(E,\beta) e^{\beta E},
$$
and therefore the configurational entropy can be obtained as,
$$
\frac{S(E)}{k_{B}} = \ln \left( W(E) \Delta E \right)
$$
where $k_{B}$ is Boltzmann's constant. Two sets of simulations
are performed to
yield monomer $\{f_{M,i}\}$ and dimer $\{f_{D,i}\}$ free energies
at the set of inverse temperatures $\{\beta_{i}\}$. Note that 
each set is known up to an arbitrary additive constant.

In the remainder of this Section, we describe a number of
transformations of the configurational partition functions $Q_{M}$ and
$Q_{D}$ either for computational convenience or for the calculation of
the proper reference state. Given the assumed integration of particle
momenta, both $Q_{M}$ and $Q_{D}$ are expressed in terms of individual
particle coordinates, and have dimensions of $V^{M}$ and $V^{2M}$
respectively, where $M$ is the number of aminoacids in the monomer.
Elimination of rigid translation or rotation degrees of freedom from the
configurational partition functions, for example, is usually
accomplished by introducing the center of mass and principal axes of
inertia of the molecule, and then relative coordinates for the
individual particles. This requires either integrals over the
corresponding canonical momenta, or the explicit consideration in the
coordinate integrals of the appropriate transformation Jacobians. Since
the Jacobians are configuration dependent, we follow instead earlier
work \cite{re:vieth95}, and conduct all our transformations on single
particle coordinates alone thus obviating the need to introduce
complicated Jacobian functions. For example, the elimination of the
degrees of freedom associated with uniform translations is accomplished
by eliminating the motion of particle 1 in one of the chains. The
elimination of rigid rotation is partially accomplished by disallowing
the rotation of the bond between particles 1 and 2 of one of the chains.
Finally, in our estimate of entropy losses on dimerization (Section
\ref{sec:reference_state}), we exclusively use phase space volumes
of single particle coordinates to maintain the necessary dimensional
consistency of Eq. (\ref{eq:kxdef}) after all the transformations of
both $Q_{M}$ and $Q_{D}$ that are described in this Section.

The configurational partition functions $Q_{M}$ and $Q_{D}$ are independent
of the location of the molecule and of its orientation. The statistical
accuracy of the simulation is greatly increased if those degrees of 
freedom that correspond to rigid translations and rotations are eliminated.
The translational degree of freedom is eliminated by fixing the
location of the first particle in the monomer chain, or of chain one
in the dimer. Since the partition function is independent of this
particle's location, this coordinate can be integrated out to yield a factor
of $V$ to the configurational partition function. The state of rigid rotation 
of the molecule can be specified by three angles, or the orientation of
one axis plus a rotation around this axis. The orientation of the axis is 
defined by its azimuth $\alpha \in (-\pi,\pi)$ and a polar angle $\theta \in
(0,\pi)$. The rotation around this axis is given by a third angle
$\gamma \in (-\pi,\pi)$. Therefore the corresponding element of volume in
configuration space is given by the triple integral
$\int d \alpha \; d \cos (\beta) \; d \gamma = 8\pi^{2}$. This value
can also be exactly factored out from the configurational partition
function. However, given that the Monte Carlo method used
employs multiple bond transitions, we have found it convenient to 
proceed somewhat differently. We disallow the 
Monte Carlo transition that corresponds to a two-bond change at the 
N-terminus, and therefore effectively eliminate the motion of 
particles 1 and 2 of the chain. While fixing the location of particle 1
still allows an exact calculation of the partition function, eliminating the 
motion of particle 2 introduces two
approximations. The first one involves the factorization of the
configuration space volume of particle 2. Since the motion 
of this particle is not
independent of the motion of the rest of the chain, this factorization
is only approximate. The degrees of freedom that are eliminated 
include a rotation around an axis defined by the bond vector between particles
one and two, plus fluctuations in bond vector
length. The elimination of the rotation is exact since the partition
function is independent of the orientation of this axis, and yields a factor
of $\int d \alpha d \cos (\beta) = 4 \pi$ to the overall
partition function. Factorization of the configuration space volume
associated with bond length fluctuations is only approximate.
We write,
\begin{equation}
\label{eq:zmono}
Q \simeq V V_{2}^{(1)} Q^{\prime} \simeq V \frac{4 \pi}{3} 
\left( R_{max}^{3} - R_{min}^{3} \right) Q^{\prime},
\end{equation}
where $Q^{\prime}$ is the partition function that is actually
computed during the Monte Carlo simulation, and $V_{2}^{(1)}$ is the 
{\em constant} accessible volume for particle 2. The second approximation
made involves the assumption that the second atom is free to move within a 
spherical shell
centered in the first atom, of inner radius $R_{min} = 4.35$ \AA ~
and outer radius $R_{max} = 7.94$ \AA. These two values are the
smallest and largest bond distances allowed in the lattice model used.
Finally, note that the computed partition function $Q^{\prime}$ still 
contains a 
factor of $2 \pi$ corresponding to the angle $\gamma$, the unrestricted rigid 
rotation of the molecule around the
axis defined by the bond vector between particles one and two that is not
eliminated during the simulation.

\subsection{Reference state calculation}
\label{sec:reference_state}

In order to introduce a common scale 
for the monomer and dimer free energy sets $\{f_{M,i}\}$ and 
$\{f_{D,i}\}$, we follow the method of ref.
\cite{re:vieth96}. At sufficiently high energies, one may assume that 
inter-chain interactions are negligible, and that the
internal motions within each dimer chain are well approximated by those
of the monomer. Therefore, and in this limit, the entropy of the dimer is 
approximately twice that of the monomer. This fact allows one to place
the entropies from both monomer and dimer 
simulations in the same reference state at high energy, and hence to 
compute free energy differences between the two.

We briefly summarize here the steps taken for both monomer and dimer. The
calculation of the internal entropy of the monomer is straightforward. In
the dimer case, however, the contributions from the
internal modes have to be separated from other
degrees of freedom related to the relative position and orientation of the
two chains. Since in the dimer simulation the two chains are
constrained to have at least one contact, the computed entropies
of the dimer at high energies still contain entropy losses due to this 
constraint that have to be estimated and subtracted
to compute its internal entropy.

\subsubsection{Monomer}

The monomer partition function $Q_{M}^{\prime}$ computed by the Monte
Carlo method still contains the contribution of one degree of freedom that
is not associated with the internal motions of the particles, and that
corresponds to a rigid rotation of the molecule around the axis defined by the
bond between particles 1 and 2. From the Monte Carlo simulation and
re-weighting, we obtain the probability density $p^{\prime}(E,\beta)$
that corresponds to the partition function $Q_{M}^{\prime}$ in Eq.
(\ref{eq:zmono}). The corresponding entropy is,
$$
\frac{S^{\prime}_{M}(E)}{k_{B}} = \ln \left( p^{\prime}(E,\beta) \Delta E
\right) + \beta E,
$$
a quantity that is independent of the inverse temperature $\beta$.
The internal entropy follows by subtracting the entropy of rotation of the 
azimuth angle of bond 2-3, and by adding
the estimate given above for the radial part of the first bond,
\begin{equation}
\label{eq:sint_m}
\frac{S_{M,int}(E)}{k_{B}} = \ln \left( p^{\prime}(E,\beta) \Delta E \right)
+ \beta E - \ln (2\pi) + \ln \frac{(R_{max}^{3}-R_{min}^{3})}{3}.
\end{equation}

\subsubsection{Dimer}

The dimer simulation is conducted by fixing the positions of atoms 1 and 2 
of one of the chains. Therefore, a Monte Carlo estimate is
obtained for $Q_{D}^{\prime}$ as given in Eq. (\ref{eq:zmono}),
As was the case for the monomer, we first define the entropy as
estimated from the simulation by
$$
\frac{S^{\prime}_{D}(E)}{k_{B}} = \ln \left( p^{\prime}(E,\beta) \Delta E
\right) + \beta E.
$$
where $p^{\prime}(E,\beta)$ corresponds to $Q_{D}^{\prime}$ above.
At sufficiently high energies, where the internal degrees of freedom
of each chain are expected to become independent of the relative
position and orientation of both chains, the total 
conformational density of states 
factors into a product involving the various contributions. In terms of
the entropy, this factorization leads to the decomposition,
\begin{eqnarray}
\label{eq:sint_d}
\frac{S_{D,int}(E)}{k_{B}} 
& = & \ln \left( p^{\prime}(E,\beta) \Delta E \right) + 
\beta E - \ln (2\pi) + \ln \frac{(R_{max}^{3}-R_{min}^{3})}{3}
- \ln V_{1}^{(2)}(E) \nonumber \\
& - & \ln \left( \varphi_{\alpha}(E) \varphi_{\cos
\beta}(E) \varphi_{\gamma}(E) \right).
\end{eqnarray}
The quantity $V_{1}^{(2)}(E)$ is the
accessible volume of particle 1 of chain 2 at energy $E$, and
hence yields the accessible volume loss of dimerization.
Its estimate during the Monte Carlo run is one of the main topics
of this paper. Since the motion of
the second chain relative to the first is constrained so that the number of 
inter-chain contacts is
greater than zero, this volume will be in general much less than $V$.
The quantity 
$\varphi_{\alpha}(E)$ is the configuration volume available for the azimuth of 
bond 1 of chain 2, $\varphi_{\cos \beta}$ for the cosine of the polar
angle of bond 1 of chain 2, and $\varphi_{\gamma}$ for the azimuth of
bond 2 of chain 2. The product of the three represents the loss of
rotational configuration space volume due to the formation of a dimer.
If the second chain were to rotate freely relative to the first, we would
have $ \varphi_{\alpha}(E) \varphi_{\cos \beta}(E) \varphi_{\gamma}(E) 
= 8 \pi^{2}$.
The value found is less that this upper bound, but
it approaches $8 \pi^{2}$ as the energy of the dimer is increased.
As was the case with $V_{2}^{(2)}(E)$,
these three quantities also need to be estimated during the simulation.

\subsubsection{Configuration space volume estimation}

The configuration space volumes $V_{1}^{(2)}(E), \varphi_{\alpha}(E), 
\varphi_{\cos \beta}(E)$ and $\varphi_{\gamma}(E)$ have been
estimated by using the method of coincidences \cite{re:ma85}. 
Let $\Gamma$ be the volume of a certain region
of configuration space. Consider a finite sample of configurations that
are uniformly distributed in $\Gamma$, and let $\Gamma_{0} \ll \Gamma$ be a 
small coarse-graining volume in configuration space. The method involves
computing the coincidence rate $R$ that a pair of configurations in the sample
belongs to the same coarse-graining volume. If the configurations are uniformly
distributed in $\Gamma$, the probability of a coincidence is $R = \Gamma_{0}
/\Gamma$. Therefore an estimate of $R = n_{c}/n_{t}$, where $n_{t}$ is the
total number of pairs in the sample, and $n_{c}$ the total number of
coincidences given $\Gamma_{0}$, allows an estimation of $\Gamma$.

In order to satisfy the conditions of the method, we first group all
configurations (regardless of their temperature) according to their energy.
Since all the configurations with the same energy are expected to occur
with equal probability we calculate the coincidence rate $R(E)$ to
estimate $\Gamma(E)$ for the various magnitudes of interest 
($V_{1}^{(2)}, \varphi_{\alpha}, \varphi_{\cos \beta},$ and 
$\varphi_{\gamma}$).

We next note some limitations in the accuracy of the method. If $\Gamma_{0}$
is not much smaller than $\Gamma$, error is introduced as $\Gamma_{0}$ will
not generate a good covering set of $\Gamma$, and it is likely that the
method will overestimate the size of the region $\Gamma$. On the other hand,
if $\Gamma_{0}$ is too small, the number of coincidences will be small,
and the statistical error in the determination of $R$ is large. There is a
third source of error associated with the sample size at each energy or,
equivalently, the total number of pairs $n_{t}(E)$ \cite{re:ma85}.
The number of coincidences can be estimated as,
$$
n_{c}(E) \sim \frac{1}{2} \left( \frac{n_{t}(E)}{k} \right)^{2} 
\frac{\Gamma_{0}}{\Gamma},
$$
with $\Gamma_{0}$ such that all $n_{t}(E)$ configurations have been 
distributed among $k$ groups. Therefore,
$$
\Gamma \sim \frac{1}{2} \left( \frac{n_{t}(E)}{k} \right)^{2} 
\frac{\Gamma_{0}}{n_{c}(E)},
$$
so that for a fixed minimum $n_{c}(E)$ to insure adequate statistics of the
coincidence rate, the estimated value of $\Gamma$ is bounded by 
$n_{t}^{2}(E)$. Therefore sufficiently large samples are needed at each
energy if the corresponding value of $\Gamma$ is large. We will further 
illustrate these limitations in Section \ref{sec:results}.

\subsubsection{Reference entropy difference}

In order to place both the monomer and dimer in the same reference
state, we require that in the limit of high $E$,
\begin{equation}
\label{eq:s0def}
S_{D,int}(E) + S_{0} = 2 S_{M,int}(E/2),
\end{equation}
where $S_{0}$ is a {\em constant}, independent of $E$, $S_{D,int}(E)$ is
given by Eq. (\ref{eq:sint_d}) and $S_{M,int}(E)$ by Eq.
(\ref{eq:sint_m}). The quantity $S_{0}$ is determined numerically as
shown in Section \ref{sec:results}.

Once the constant $S_{0}$ has been determined, the free
energy and partition function of the dimer are re-scaled according to
\begin{equation}
\label{eq:zredef}
f_{i}^{\prime\prime} = f_{i}^{\prime} - S_{0}, \quad
Q^{\prime\prime}_{D} = Q^{\prime}_{D} e^{S_{0}}.
\end{equation}

We can now compute the equilibrium constant $K_{x}$ by
substituting Eq. (\ref{eq:zmono}) for both monomer and dimer into 
Eq. (\ref{eq:kxdef}), but using the rescaled
dimer partition function $Q_{D}^{\prime \prime}$ defined in 
Eq. (\ref{eq:zredef}) instead of $Q_{D}^{\prime}$,
\begin{equation}
\label{eq:kx3}
K_{x} = \frac{N}{V V_{2}^{(1)}} \frac{Q^{\prime\prime}_{D}}{\sigma_{D}
Q_{M}^{\prime \; 2}} = \frac{N}{V V_{2}^{(1)}}
\frac{e^{S_{0}} Q^{\prime}_{D}}{\sigma_{D} Q_{M}^{\prime \; 2}}.
\end{equation}
With this re-definition of the dimer partition function, both 
$Q_{D}^{\prime\prime}$ and $Q_{M}^{\prime}$ are referred to the same
reference state, and hence absolute values of $K_{x}$ can be given.

In terms of the free energies of the monomer and dimer that are obtained 
from the Monte Carlo calculation after re-weighting (Eqs. 
(\ref{eq:probability_def}) and (\ref{eq:freeenergy_def})), Eq.
(\ref{eq:kx3}) leads to,
\begin{equation}
\ln K_{x}(\beta_{i}) = \ln c_{0} + S_{0} -f_{D,i}^{\prime} - \ln \sigma_{D} - 
\ln V_{2}^{(1)} + 2 f_{M,i}^{\prime}.
\end{equation}
If $\ln K_{x} > 0$ the dimer is prevalent. 

\section{Results}
\label{sec:results}

The method described in Section \ref{sec:model} has been tested on the
GCN4 leucine zipper (a 31 residue segment with the characteristic
heptad repeat sequence of leucine zippers). The oligomerization
equilibrium of the wild type has been addressed both experimentally
\cite{re:oshea91} and computationally \cite{re:mohanty99}, as well as
that of several of its mutant forms \cite{re:harbury93,re:vieth95}. Due
to the short length of the sequence, and the simplicity of its secondary
structure, numerous computational studies have addressed various aspects
of the oligomerization process in GCN4, including dimer and multi-mer
equilibria \cite{re:vieth95}, the stability of several of its sub-domains
\cite{re:vieth96}, oligomeric equilibrium of several of its mutant forms
\cite{re:skolnick95}, and other parameters of the coiled coil such as
the helical content as a function of temperature and a van't Hoff enthalpy 
analysis to reveal the adequacy of a two state assumption for the 
dimerization process \cite{re:mohanty99}.

We have extended the analysis of \cite{re:mohanty99} in two directions. First
we use a Replica Exchange Monte Carlo method instead of the Entropy 
Sampling Monte Carlo method of that reference as the former provides a faster
rate of convergence to the equilibrium distribution of the dimer form.
Second, we extend the method of calculation of the various entropy losses 
upon dimerization, and show their strong dependence on the energy of the
configuration, a dependence that was not taken into account in
previous studies.

\begin{figure}
\vspace{2cm}
\epsfig{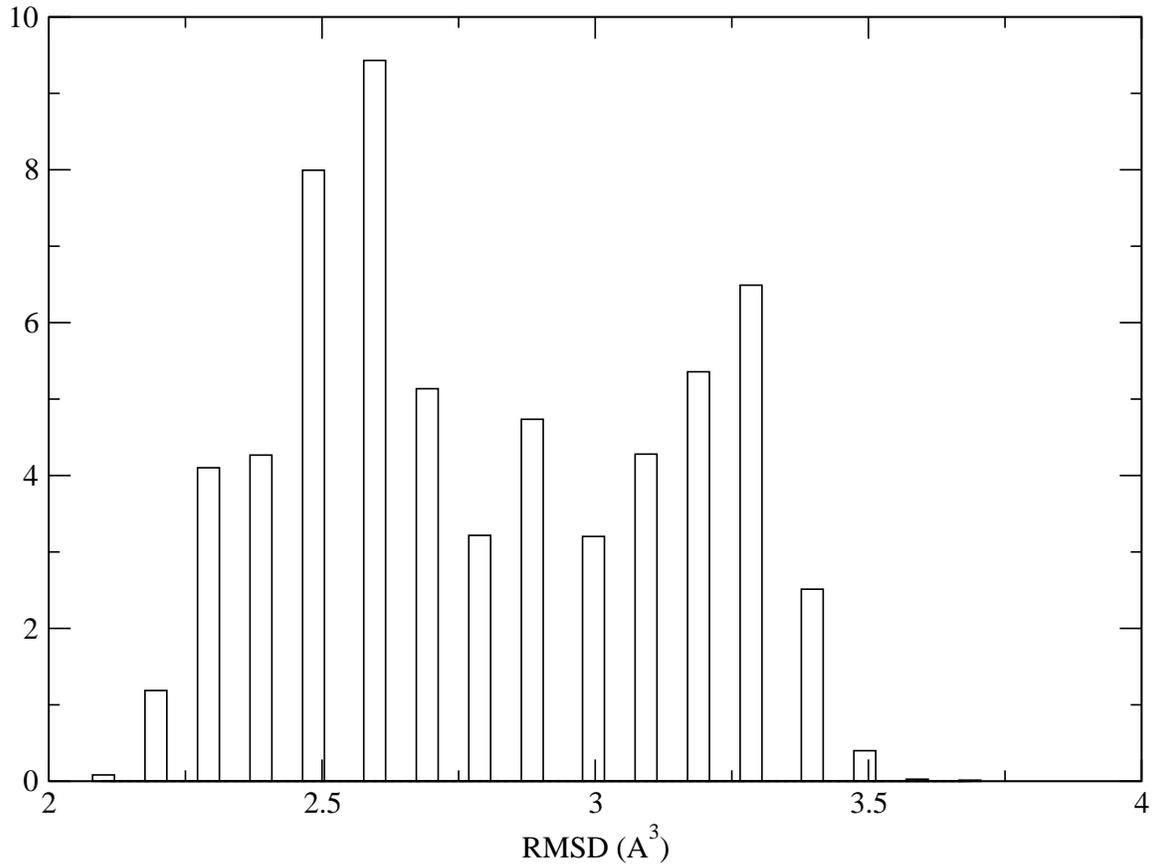}
\vspace{0.5cm}
\caption{Histogram of the sampled root mean squared deviation from
native (RMSD in \AA$^{3}$ at the dimensionless temperature $T = 0.6$.)
.}
\label{fi:rmsd}
\end{figure}

The results shown are based on two long runs for the monomer and
dimer forms respectively. Initial configurations were chosen close to the
native state, but first equilibrated at constant temperature. 
Several runs with different initial conditions yielded essentially identical
results for the various thermodynamic quantities presented, although none
of the dimer simulations involved an initial condition in a manifestly 
anti-parallel configuration. The monomer runs involved
$2\times10^{6}$ independent configurations or steps after equilibration, with
one replica exchange attempted every 500 steps. Quantities for analysis
were collected every 250 steps. The dimer
simulation comprises two identical, and initially parallel chains with
at least one contact between them \cite{fo:js1_1}.
The simulation in this case is conducted by rejecting all bond moves that
would result in no contacts between the chains.  The run for the dimer involved
$1.3\times10^{6}$ configurations, with the same frequency of analysis and
of replica exchange. In both cases, twenty independent replicas were run in 
parallel at
dimensionless temperatures in the range $T = 0.5 - 1.45$ in increments
of 0.05. In the low temperature range, the root mean squared deviation
(RMSD) between the 
estimated location of the $\alpha$ carbons in the model and the
native configuration is of the order of 3 \AA (see Fig. \ref{fi:rmsd}). We 
note that this
RMSD range constitutes a prediction, and was not enforced during
the course of the simulation.

The analysis presented is based on energy histograms, and the
subsequent re-weighting described in Section \ref{sec:model}. The entire
range of energies sampled by the monomer and the dimer during the course
of the simulation was divided into 100 equal bins, so that in
dimensionless units $\Delta E \simeq 0.2926$ for the monomer and
$\Delta E \simeq 0.3767$ for the dimer.

In the case of the dimer, the location of all individual particles was
also recorded every 250 steps in order to estimate the configuration
space volumes $V_{1}^{(2)}, \varphi_{\alpha}, \varphi_{\cos \beta}$ and
$\varphi_{\gamma}$. The results presented for $V_{1}^{(2)}$ are based on
the spatial coordinates of particle 15 of chain 2 (the chain that 
is free to move within the computational cell). Substantially
identical results follows from an analysis of any other particles in
the chain, except for those in the immediate vicinity of the N- or
C- termini.

The configuration space volume $\varphi_{\alpha}$ is estimated from the
azimuth of the bond between particles 15 and 16 of chain 2, 
and $\varphi_{\cos \beta}$ follow from $\cos \beta$, $\beta$ being 
the polar angle of this bond. Finally, $\varphi_{\gamma}$ is
obtained from the azimuth distribution of the bond between particles
15 and 16. In a freely rotating molecule, $\alpha$ is uniformly
distributed in $(-\pi,\pi)$, $\cos \beta$ in $(-1,1)$, and
$\gamma$ in $(-\pi,\pi)$, resulting in a combined conformational space
volume for rigid rotation of $\varphi_{\alpha} \varphi_{\cos \beta}
\varphi_{\gamma} = 8 \pi^{2}$.

\begin{figure}
\vspace{2cm}
\epsfig{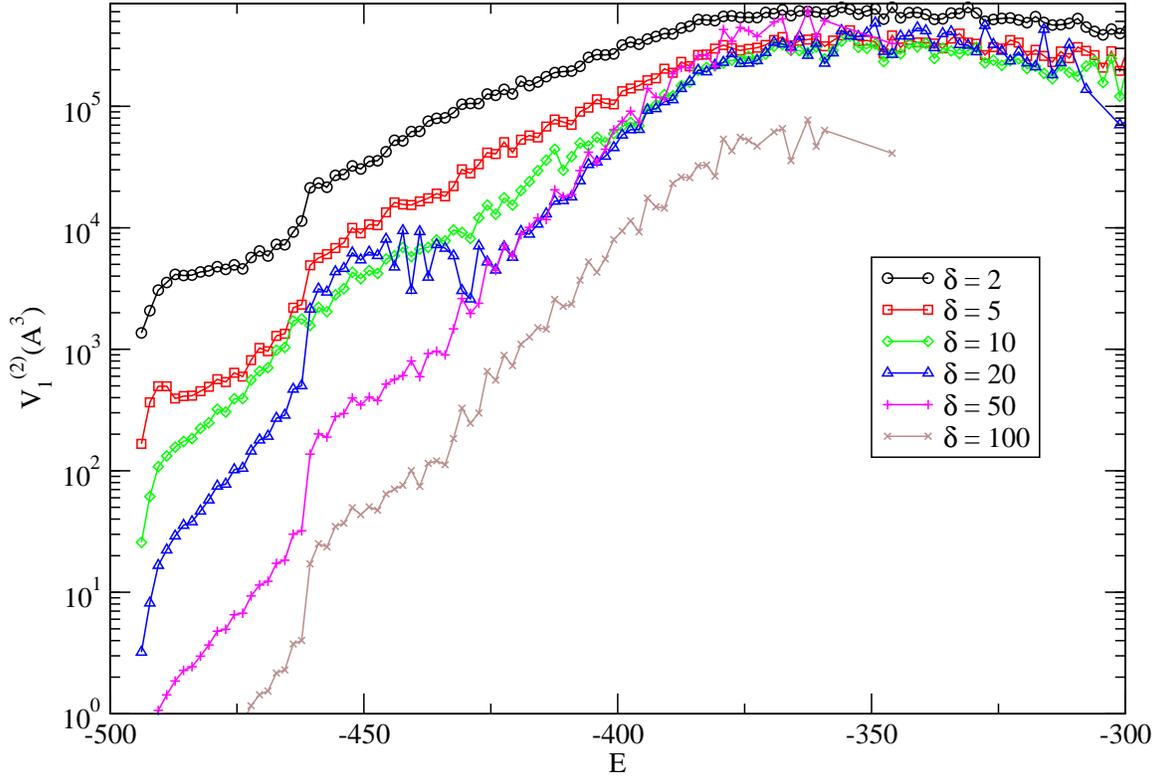}
\vspace{0.5cm}
\caption{Volume of configuration space $V_{1}^{(2)}$ (in \AA$^{3}$) 
that is accessible 
to rigid translation of chain 2 of the dimer as a function of its energy.
Several different choices 
of the coarse-graining volume $\Gamma_{0}$ are shown in the figure as 
indicated by the values of $\delta$.}
\label{fi:scm}
\end{figure}

Figure \ref{fi:scm} shows our results for $V_{1}^{(2)}$ with the same
energy bin size $\Delta E$ used to construct the histogram. The 
coarse-graining volume $\Gamma_{0} = \Delta x \Delta y \Delta z$ has
been obtained by defining $\Delta x = (x_{max} - x_{min})/\delta$,
and similarly for $\Delta y$ and $\Delta z$. $x_{min}$ and $x_{max}$
are the smallest and largest values of $x_{15}$, the $x$ coordinate
of particle 15, for each particular energy bin. We present our
results for a range of values of $\delta$ in Fig. \ref{fi:scm}. If
$\delta$ is too large, the coarse-graining volume is small, and the number
of configurations for a given energy $n_{t}(E)$ is also small. As 
discussed in Section \ref{sec:model}, this leads to underestimate
the accessible volume. The value of $V_{1}^{(2)}$ is seen to increase
with decreasing $\delta$, becomes approximately independent of $\delta$ in
some range, and then further increases with decreasing $\delta$. If $\delta$
is too small, the shape of the region being sampled cannot be 
accurately reproduced with this coarse $\Gamma_{0}$. Note that the
values of $V_{1}^{(2)}$ at low energies are the most difficult to
estimate, presumably because the shape of the region in configuration
space is not as smooth as that at higher energies. However since the
procedure leading to the computation of the reference entropy relies
only on the region of high energies, this inaccuracy does not represent a
significant limitation to our results.

\begin{figure}
\epsfig{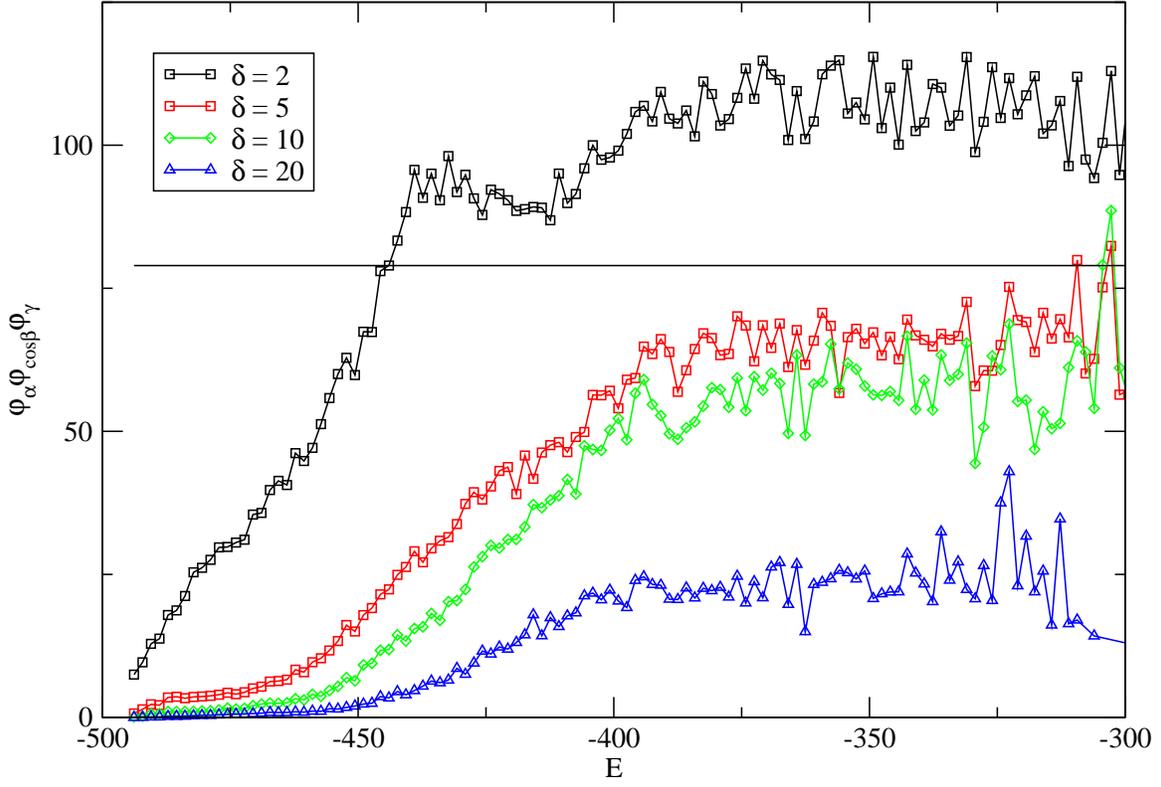}
\vspace{0.5cm}
\caption{Volume of configuration space accessible to rotation of
chain 2 of the dimer $\varphi_{\alpha} \varphi_{\cos \beta} 
\varphi_{\gamma}$ for four different choices of the coarse-graining
volume, as indicated by the values of $\delta$ in the figure. The
straight line corresponds to the value associated with free rotation, 
$8 \pi^{2}$.}
\label{fi:sangle}
\end{figure}

The behavior just described is qualitatively similar to that shown in
Fig. \ref{fi:sangle} corresponding to the rotation volume
$\varphi_{\alpha} \varphi_{\cos \beta} \varphi_{\gamma}$. In this
case we define $\Gamma_{0} = \Delta \alpha \Delta (\cos \beta) \Delta 
\gamma$ with the same definition of $\Delta \alpha, \Delta (\cos \beta)$ and
$\Delta \gamma$ in terms of the quantity $\delta$. The figure also shows 
(solid line) the value $8\pi^{2}$ that corresponds to free rotation of
chain 2 relative to chain 1. As can be seen from the figure, the values
obtained approach this limit at high energies.

\begin{figure}
\epsfig{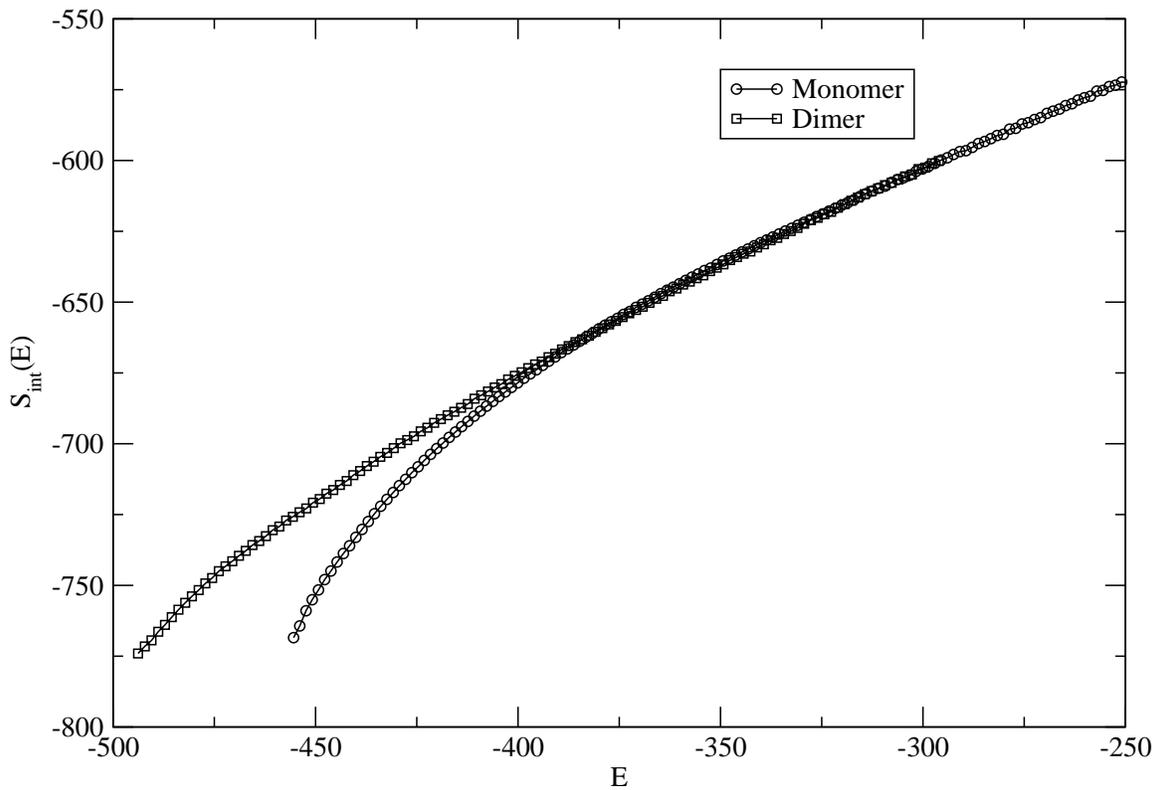}
\vspace{0.5cm}
\caption{Rescaled internal entropy of the dimer $S_{D,int} + S_{0}$,
with $S_{0} = 52$ (squares), and twice the internal entropy of the monomer
$2S_{M,int}(E/2)$ (circles). The value of $S_{0}$ has been
graphically determined in order to make the two curves coincide 
in the range of large $E$. As can be seen from the figure, the two curves 
superimpose quite accurately for a significant range of energies.}
\label{fi:s0}
\end{figure}

The constant $S_{0}$ of Eq. (\ref{eq:s0def}) required to place both the 
monomer and dimer free energies in the same scale is obtained directly
from Eqs. (\ref{eq:sint_m}) and (\ref{eq:sint_d}), as shown in Fig.
\ref{fi:s0}. In this figure we plot $S_{D,int}(E) + S_{0}$ and
$2S_{M,int}(E/2)$ with $S_{0}$ adjusted graphically so that the two 
curves coincide at large $E$. Note that both curves superimpose
to a good accuracy for a range of energies, indicating the 
consistency of the approach.

\begin{figure}
\epsfig{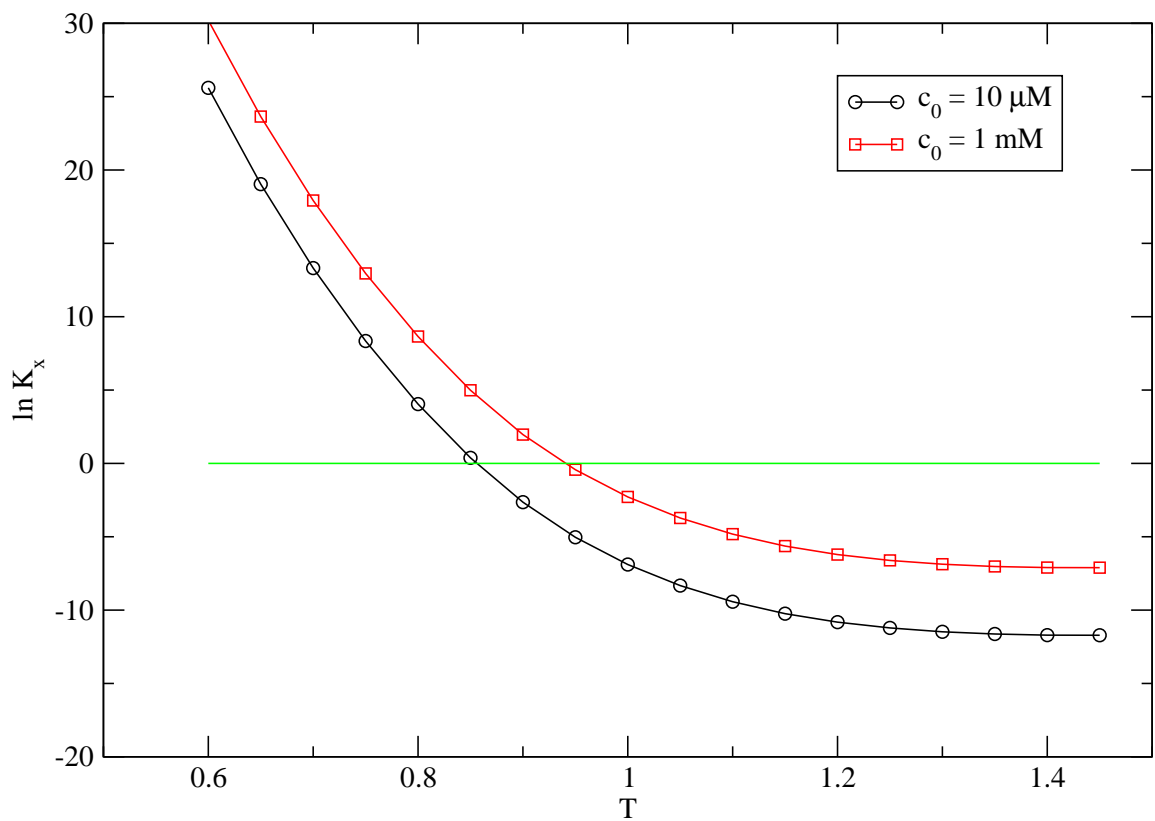}
\vspace{0.5cm}
\caption{Equilibrium reaction constant for monomer-dimer equilibrium as a 
function of dimensionless temperature and for the two concentrations 
indicated.}
\label{fi:kx}
\end{figure}

We show next our results for $K_{x}$ in Fig. \ref{fi:kx}, with
$K_{x}$ defined in Eq. (\ref{eq:kx3}), as a function of the
dimensionless temperature. When $K_{x}=1$ the mol fractions
of the monomer and dimer forms are equal ($x_{M} = x_{D} = 0.5$).
At low temperatures $K_{x} > 1$, indicating
a prevalence of the dimer form, and the reverse is true at high
temperatures.  For
the sake of illustration, the figure shows the values of $K_{x}$ at
two different concentrations $c_{0} = 10~\mu M$ and $c_{0} = 1~mM$.

\begin{figure}
\epsfig{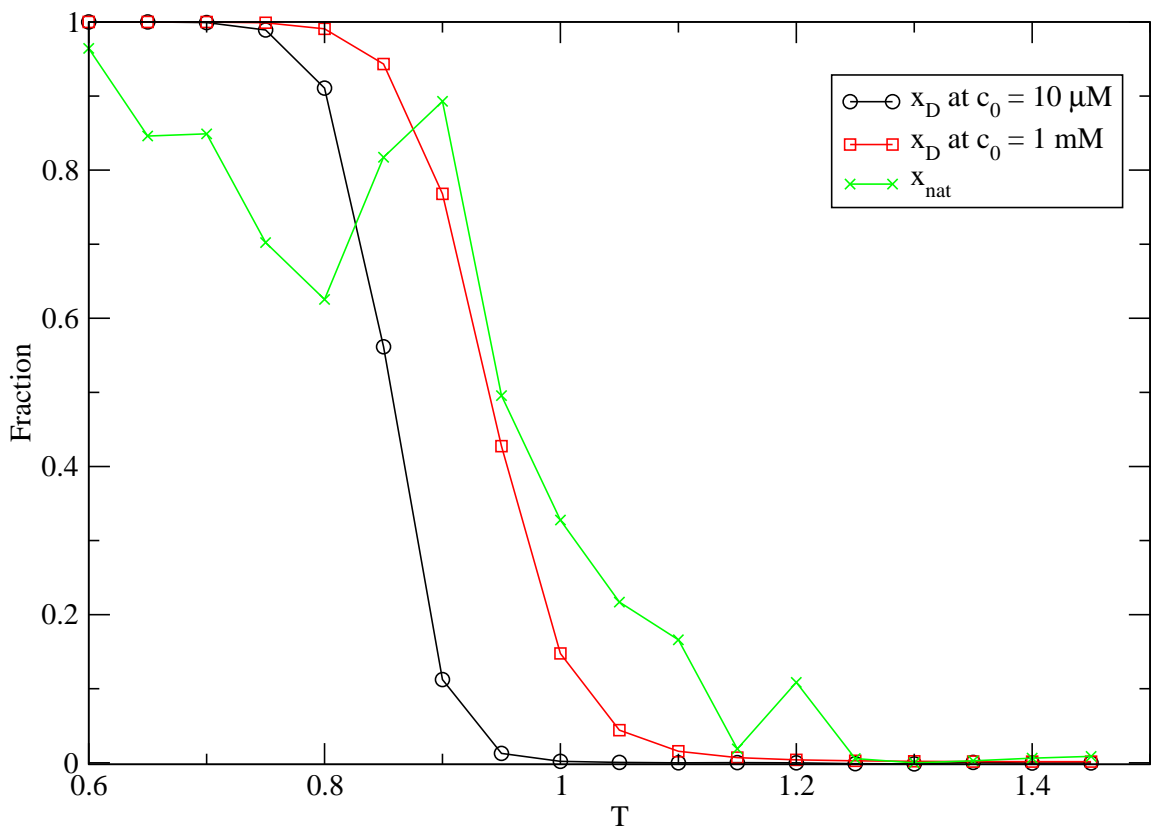}
\vspace{0.5cm}
\caption{Equilibrium mol fraction of the dimer as a function of
temperature. $x_{D} = \left( 1 + 4 K_{x} - \sqrt{1 + 8 K_{x}} \right) /
4 K_{x}$, for the two concentrations indicated. The value of the
equilibrium constant $K_{x}$ is shown in Fig. \protect\ref{fi:kx}. We
also show the fraction of configurations that, at the given temperature,
had at least 50\% native contacts.}
\label{fi:cmap}
\end{figure}

It is also intersting to examine the contact map of the dimer phase
as given by the simulation. As discussed above, the only constraint in
the simulation is that there be at least one contact between the two
chains. Therefore the question arises as to whether the dimer retains a
significant fraction of native contacts in the vicinity of the transition 
temperature, or whether there is a
significant fraction of out of register dimer configurations that are 
structurally very different from the native state. In order to answer
this question, we proceed as follows: A
contact between residues belonging to different chains is considered native 
if it appears in the contact map of the native protein in its lattice 
representation. With this definition, GCN4 has 10 inter chain native contacts.
We then calculate the ensemble average of the fraction of
configurations that have {\em at least} 50 \% native contacts.
The results are shown in Fig. \ref{fi:cmap} as a function of
temperature. This fraction approaches one at low temperature, changes 
quickly around the transition region, decaying to zero at high
tempertures. We also shown in this figure the equilibrium mol fraction of the 
dimer form for the same two concentrations shown in Fig. \ref{fi:kx}.
From the figure, we conclude that in this range of physiological
concentrations, the decrease in the fraction of native contacts as given 
by our model can be mainly attributed to the appearance of the monomer form, 
and not to a significant contribution from out of register dimers.

To conclude, our results show that it is possible to calculate the entropy
loss of dimerization corresponding to the GCN4 monomer-dimer equilibrium
without any restrictions to the motion of the individual chains. Previous
research on this system differed from ours in that the entropy
loss was estimated by restricting the conformation space of the dimer,
thus resulting in low values of the entropy 
(compare the value of $V_{1}^{(2)} = 67.6$\AA$^{3}$ given in Table 1 of ref. 
\cite{re:vieth96}, and the values shown in Fig. \ref{fi:scm}).
Despite the fact that we have allowed sampling of the full conformational 
space of the dimer, the results obtained confirm that it is possible to 
obtain the free energy of dimerization of GCN4 at physiological
concentrations by using a reduced model of the protein and Monte Carlo
simulations.

\section*{Acknowledgments}
This research has been supported by the National Science Foundation,
grant No. 9986019. JV is also supported by the National Institutes of 
General Medical Sciences, grant No. GM64150-01.

\bibliographystyle{prsty}
\bibliography{$HOME/mss/references}

\begin{thebibliography}{10}
\bibitem{fo:js1_1}
Two residues are considered to be in contact if the distance between the two
  corresponding particles is smaller than a predefined cut-off. The value of
  this cut-off distance depends on the pair of aminoacids involved, and has
  been determined simultaneously with the other parameters that define the
  interaction potentials. Distances range between approximately 3 and 5 in
  units of the lattice spacing.
\end{thebibliography}

\newpage
\appendix
\section{Lattice protein model and interaction force parameters}
\label{ap:sicho}

The model protein used in this work employs a reduced representation of
the protein backbone on a regular lattice. The model comprises a 
sequence of bonds connecting particles located at the center of mass
of the corresponding residue and backbone $\alpha$ carbon. The
particles are then placed in a three dimensional simple cubic lattice with
spacing of 1.45 \AA. Further details on this model can be found in
ref. \cite{re:kolinski99}

A sequence of configurations is generated by a Monte Carlo scheme
with Metropolis updating. The method employs three different types of
individual transitions. In the first case, a single particle and its
two corresponding bonds are selected for an attempted update. The second
type of transition involves three consecutive bonds and the corresponding
two adjacent particles. The third involves a rigid translation of a small
fragment of the chain comprising three particles, and the ensuing
rearrangement of the end bonds. These transitions are attempted
sequentially for all the bonds in the chain. Two separate transitions are also
included to adjust the position of the N- and C-termini particles. 
The set of all these attempted transitions constitutes a Monte Carlo step
(MCS). Further details about the transitions used in the Monte Carlo updating 
can be found in ref. \cite{re:kolinski98b}.

Interaction forces can be grouped into generic and sequence specific.
The former are sequence independent and lead to protein-like packing, whereas 
the latter are derived from a statistical analysis of the protein database, 
and explicitly depend on the identity of the aminoacids involved.
We next list the values of the various parameters used in our
calculations. Sequence dependent short range interactions are defined by
Eq. (1) of \cite{re:kolinski98b}. We use a common multiplicative factor in our
calculations $\varepsilon_{short}=0.325$ (this factor is explicitly shown
in Eq. (12) of \cite{re:kolinski99} with a value of 0.75 instead). A
three-body potential that is sequence specific is also used, with an
amplitude $\varepsilon_{3b}=0.25$. The generic, short range
conformational biases of Eqs. (3), (4), and (5) of \cite{re:kolinski99}
involve $\varepsilon_{gen} = 1.25$. Hydrogen bonding energies within the
main chain are also included, with an amplitude $\varepsilon_{H-bond} =
0.325$ in Eq. (7) of \cite{re:kolinski99}. Long range and sequence
dependent interactions are modeled by a set of square well potentials as
described in Eq. (9) of \cite{re:kolinski99}. We have chosen $E^{rep} =
4$ in Eq. (8) of that reference, and a common multiplicative factor 
$\varepsilon_{pair} = 2.0$ (to be compared with the factor of 1.25 in
Eq. (12) of \cite{re:kolinski99}). Two additional multibody potentials
are introduced to include hydrophobic effects, and preferences for
parallel or anti-parallel packing among the residues. We define as the
scale of Eq. (10) in \cite{re:kolinski99} $\varepsilon_{surface}=0.75$
(instead of the value 0.5 shown in Eq. (12) of that reference). Finally,
we have used a factor $\varepsilon_{multi}=0.75$ in Eq. (11) of
\cite{re:kolinski99} (instead of the value 0.5 in Eq. (12)).

\newpage

\newpage

\newpage

\newpage

\newpage

\newpage

\end{document}